\documentstyle[twocolumn,prl,aps]{revtex}
\large
\newcommand{\be}{\begin{equation}}
\newcommand{\ba}{\begin{array}}
\newcommand{\ea}{\end{array}}
\newcommand{\beq}{\begin{eqnarray}}
\newcommand{\eeq}{\end{eqnarray}}
\newcommand{\ee}{\end{equation}}

\def\w{\omega}

\def\l{\lambda}

\begin{document}
\draft
\twocolumn[
\hsize\textwidth\columnwidth\hsize\csname @twocolumnfalse\endcsname

\title{
Temperature-dependent vibrational heterogeneities in harmonic
glasses}
\author{G. Viliani }
\address{Dipartimento di Fisica and INFM, Universita di Trento, I-38050
Povo, Trento, Italy}
\author{E. Duval}
\address{Laboratoire de Physicochimie des Mat\'{e}riaux Luminescents,
Universit\'{e} Lyon I - UMR-CNRS 5620  43, boulevard du 11 Novembre 69622
Villeurbanne Cedex, France}
\author{L. Angelani}
\address{Dipartimento di Fisica, INFM and INFM - Center for Statistical Mechanics and Complexity, 
Universita' di Roma "La Sapienza",
00185 Roma, Italy}

\date{\today}
\maketitle
\begin{abstract}
Numerical simulation is employed to study dynamical heterogeneities in model
harmonic glasses whose atoms interact via three variants of the
Lennard-Jones potential
(monoatomic full Lennard-Jones, soft spheres, binary mixture).
Heterogeneities are observed to exist in all three kinds of glasses, and in
some cases
they are observed to depend on temperature. The dimension of the
heterogeneities is studied for the full Lennard-Jones case.
\end{abstract}
]

The question as to whether the structure of glasses at the scale of tens to
hundreds of interatomic distances is homogeneous or inhomogeneous has
attracted recently much interest. Traditionally, following the continuous
random network model suggested by Zachariasen \cite{zach32}, the homogeneous
hypothesis has prevailed, also because no heterogeneity was clearly observed
by small angle neutrron or X-ray scattering and electron microscopy.
However, these techniques show only that there is no evident density
heterogeneity, but tell nothing about the cohesion or elasticity at the
nanometric scale, and thus cannot rule out vibrational {\it dynamical}
heterogeneities.

The inhomogeneous cohesion of glasses at the nanometric scale was invoked
to interpret the excess of vibrational density of states  in the terahertz
frequency range, the so-called boson peak \cite{duv90,duv98}. By this
interpretation, the frequency of the
boson peak is related to the inhomogeneity size.

Actually, recent molecular dynamics simulations on binary Lennard-Jones (LJ)
mixtures \cite{Kob97,Poo98,Don98} showed that in the supercooled state
highly mobile and immobile particles are spatially correlated over a range
that grows with temperature as the glass transition is approached; this
finding is  consistent with the observation of dynamical heterogeneities in
supercooled liquids \cite{Sch91,Tra98}. One important issue is whether or
not such heterogeneities are in some sense "freezed down" through the glass
transition, so that even in the cold, harmonic glass there exists a memory
of these heterogeneities and, as a consequence, softer and harder zones.

In another recent simulation this problem was addressed by calculating the
local pressure or stress in a glassy LJ system \cite{kus00}, and a strong
correlation was found among atoms experiencing the same local
pressure. In this reference, the elastic constants pertinent to one atom
(which are related to local pressure) were determined by keeping all the
other atoms fixed \cite{kus00}, and if the average local elastic constant
was large (small) the atom was ranked as hard (soft). Finally, very recently
the vibrational modes of 2-D LJ clusters were determined numerically
\cite{Wit01}; it was found that at length scales less than approximately 40
atomic distances the elasticity is no more affine with respect the sample
deformation, and that heterogeneities of force constants are extended on the
same nanometric lengths \cite{Wit01}.

In the present paper we investigate the possible effect of temperature on
the
heterogeneities which are found in {\it harmonic} glasses of the LJ family.
The existence of such effects is expected in principle
on the basis of the following
argument: Consider a glass at low temperature; in this case only the modes
of low frequency will be excited and the atoms will only perform the
cooperative motions corresponding to the normal modes in question;
therefore, irrespective of the magnitude of the {\it local} force constants,
motion in a given direction will only occur if allowed by the excitable
modes. In this sense, the heterogeneities observed in Ref. \cite{kus00} can
describe the high-temperature behaviour, and otherwise it is expected
that if disomogeneities exist, their shape, dimension,
quantity etc. may depend on temperature.

We considered monoatomic systems interacting through the LJ potential
($N_0=$2048 atoms) or the repulsive part of the LJ potential (soft spheres,
$N_0=$2048),
and the bynary LJ mixture introduced in Refs. \cite{Kob97,Poo98,Don98}
($N_0=$2000). The potential parameters, mass and
density of the LJ and soft-sphere
system were those suitable for Argon \cite{frustra};
for the binary mixture we chose the
potential parameters as in Refs. \cite{Kob97,Poo98,Don98}. The stable
configurations were produced by a fast quench (down to typically $T_m/10$,
where $T_m$ is the melting temperature) of the liquid configurations
starting from temperatures just above $T_m$. The liquid configurations were
obtained by standard MD equilibration runs. The frozen configurations at
$T_m/10$ were then further equilibrated in order to check against
crystallization. Finally, a steepest descent was applied to find the
(glassy) minimum configuration at $T=0$. For all systems we averaged the
quantities of interest over 5 different realizations.

Following the procedure of Ref. \cite{frustra}, soft (hard) atoms, relative
to a given eigenvector, are identified as those with small (large) variation
of potential energy under the displacement pattern produced by the
eigenvector itself. Therefore, in the harmonic approximation, the dynamical
matrix was diagonalized, thus obtaining eigenvalues and eigenvectors.
Subsequently, for each atom $i$ and for each normal mode $p$, we
calculated the potential-energy change due to the collective motion
determined by the normal mode in question:
$$
        \delta V_i^p(T) = \sum_{j\neq i} V(r^p_{ij}(T)) - V(0)
$$
Here the sum runs on all atoms other than $i$, $r^p_{ij}(T)$ is the set of
atomic distances resulting from the displacements produced by normal mode
$p$, whose amplitude depends on temperature; $V(0)$ is the potential energy
of the equilibrium configuration. At a given temperature, all normal modes
are excited, with amplitude depending on their energies $\hbar \w_p$. Since
the motions due to the different modes are not mutually coherent, we
assume that the
total potential energy variation is the sum of the weighted
variations produced by the single modes:
\be
       \delta V_i(T) = \sum_p \delta V_i^p(T)
\ee
The distances $r^p_{ij}(T)$ can be written as $[r_i^p(T)-r_j^p(T)]$, where
the atomic positions depend on the eigenvector displacements:
$$
        r_i^p(T)=r_i^0 + Q^p(T)e_i^p,
$$
and the amplitudes $Q^p$ are the temperature-dependent quantities. For a
normal mode of frequency $\w_p$, and at temperature $T$, the average
vibrational level is
$$
                 <n^p(\w_p,T)>=\frac{1}{[exp(\hbar\w_p/kT) - 1]},
$$
and the expectation value of $Q^2$ is calculated in the state $|n>$ as:
$$
      Q^2 = <n|Q^2|n> = \hbar/(m\w_p)(2n+1),
$$
and consequently $Q^p(T) = \sqrt{Q^2}$ is determined as a function of $T$
and $\w_p$, and the temperature-dependent potential-energy variations are
computed through Eq. (1).
In what follows, temperatures are alway expressed in units of the
maximum frequency $\w_{M}$ of the respective systems, $T=\frac{\hbar
\w_{M}}{k} $.

The presence of vibrational  heterogeneities is detected by comparing the
normalized pair correlation function of the whole system $g_W(R)$ to that of
the $N$ softest and/or hardest atoms, $g_N(R)$ \cite{kus00}; in fact, for
example, if the nearest-neighbor or next-nearest-neighbor peaks in $g_N$ are
higher than in $g_W$ it means that the $N$ softer or harder atoms tend to
group.

In Fig. 1 we report the pair correlation functions $g_N(R)$ of the
monoatomic full LJ system for the $N=20, 50, 100$
and $200$ softest atoms at $T=0.3$, and $g_W(R)$ (which is $T$-independent).
Clear differences appear in the first peak: that of all selected groups of
soft atoms is narrower, more intense than for the whole, and shifted to
shorter distances, indicating that the soft atoms belong to the
low-$R$ tail. Excess correlation is observed on the next-neighbour peaks,
but only for the smallest value $N=20$, whose $g_N(R)$ is very noisy: however,
the difference is so marked that such peaks might be significant.

In Fig. 2 is shown, for the same system,
the temperature variation of the nearest-neighbour peak
of $g_N$ for $N=100$, for soft ($a$) and hard ($b$) atoms. A marked
temperature dependence is observed for the soft ones, together with a jump
in correlation between $T=0.1$ and $T=0.3$. For both kinds of atoms, the peak
is shifted to low $R$ values, but for the soft ones it becomes narrower than
in $g_W$ at high $T$; this can be understood by considering that the modes
of low frequency, i.e. of long wavelength, are weakly dependent on the local
variation of the bonding between neighbouring atoms, the dependence
increases with the frequency of the modes so that the softest atoms become
better differentiated from the whole. The hard atoms show little or no
temperature variation on the highest point, but contrary to soft ones 
the overall area under the nearest-neighbour peak is increased
with respect to $g_W$, especially at the two highest $T$-values.

The nearest-neighbour peaks of soft spheres are reported in Fig. 3. In this
case again there is very little $T$-variation, but the excess
correlation is
quite marked across the whole peak whose area is much larger than in $g_W$;
the low-$R$ side of hard atoms seems again to be favoured.

In Fig. 4 are reported the nearest-neighbour peaks for the LJ binary
mixture. The major effects concern the low-$R$ tail of the $A-A$ peak
and the whole $A-B$ peak. For soft atoms, there is a strong, $T$-dependent
increase of the former and a corresponding, $T$-dependent decrease of
the latter. The same, but less $T$-independent, changes occur
for hard atoms.

Summarizing the results presented so far, it can be said that for the three
harmonic systems studied dynamical heterogeneities are clearly observed;
the most pronounced overall correlation increases
 are for the binary mixture and for soft spheres,
while soft inhomogeneities are observed to depend on temperature for
the full LJ and binary mixture. In all cases, the most pronounced effects
concern the low-$R$ tail of the pair correlation function.

One important question is whether with the present system size ($\approx$
2000 atoms) it is possible to estimate the expected size, $\l$,
of the vibrational dynamical inhomogeneities. To check this point, in the
case of
full LJ we evaluated $\l$ for samples with different number of atoms,
$N_0=500, 1000, 1500, 2048$.
In order to evaluate $\l$, the height of the first point of $g_N(R)$ (marked
by an arrow in Fig. 2) was evaluated as a function of $N$ and compared
to the corresponding height in $g_W$; the value of $N$ at which
the two become equal is taken as an estimate of the number of atoms which
constitute the inhomogeneity and hence of its size. The results are
reported in Fig. 5 and show that certainly for hard atoms, and
probably for soft ones as well at high temperature,
the heterogeneity size is still
growing with $N_0$ at the maximum sample dimension available to us, which
seems to indicate that the "true" size is larger than we can determine.
One possible exception to this is the low-temparature ($T=$0.01 and 0.1)
soft atoms, for
which there appears to be a flattening of the curve in Fig. 5 starting at
$N_0$=1000. However, before drawing definite conclusions we think it is
wiser to await for more statistics (only 1 sample was examined for
$N_0=$500 and 1000, and 3 samples for $N_0$=1500). However,
if the low-$T$ result of Fig. 5 
should be confirmed, temperature would play a role also in determining
the soft heterogeneity size, and the same should be expected for the
binary mixture (see Fig. 4(a) ).

In summary, it is the first time that  the existence of $T$-dependent
vibrational
dynamical heterogeneities in glasses is clearly shown to exist by
simulation with LJ or soft spheres systems. However, it is not possible to
draw definite and general conclusions
about the heterogeneity size and its relation with the boson peak
\cite{duv90,duv98}. For that, the number of atoms in the samples studied
by simulation should be much larger than 2000; work in this direction is
in progress. We also plan
to use other potentials taking into account long-range
interactions, like for example the BKS one \cite{Bee90} which is suitable for
SiO$_2$. We expect a more
contrasted inhomogeneous elasticity or vibrational inhomogeneity, like in
vitreous silica, possibly in relation with a more intense boson
peak \cite{Jun01}.

\newpage

\begin{figure}
\label{f1}
\caption{Pair correlation function $g(R)$, in a monoatomic system, simulated
with the full Lennard-Jones potential, for the 20, 50, 100 and 200 softest
atoms
compared to $g(R)$ of the whole, at $T$=0.3. $L$ is the size of the cubic
simulation box.}
\end{figure}

\begin{figure}
\label{f2}
\caption{Nearest-neighbour peak of $g(R)$ for the 100 softest ($a$) and
hardest ($b$)
atoms of the full Lennard-Jones system, at various temperatures. Arrows:
see text.}
\end{figure}

\begin{figure}
\label{f3}
\caption{Same as Fig. 2, but for soft spheres (repulsive part of
Lennard-Jones).}
\end{figure}

\begin{figure}
\label{f4}
\caption{Same as Fig. 2, but for the binary $A-B$ (80\%-20\% respective
concentration)
mixture introduced in Ref. 4. The higher peak corresponds to
$A-A$ neighbours,
the lower one to $A-B$ ones. The $B-B$ peak is quite low and masked between
the two.}
\end{figure}

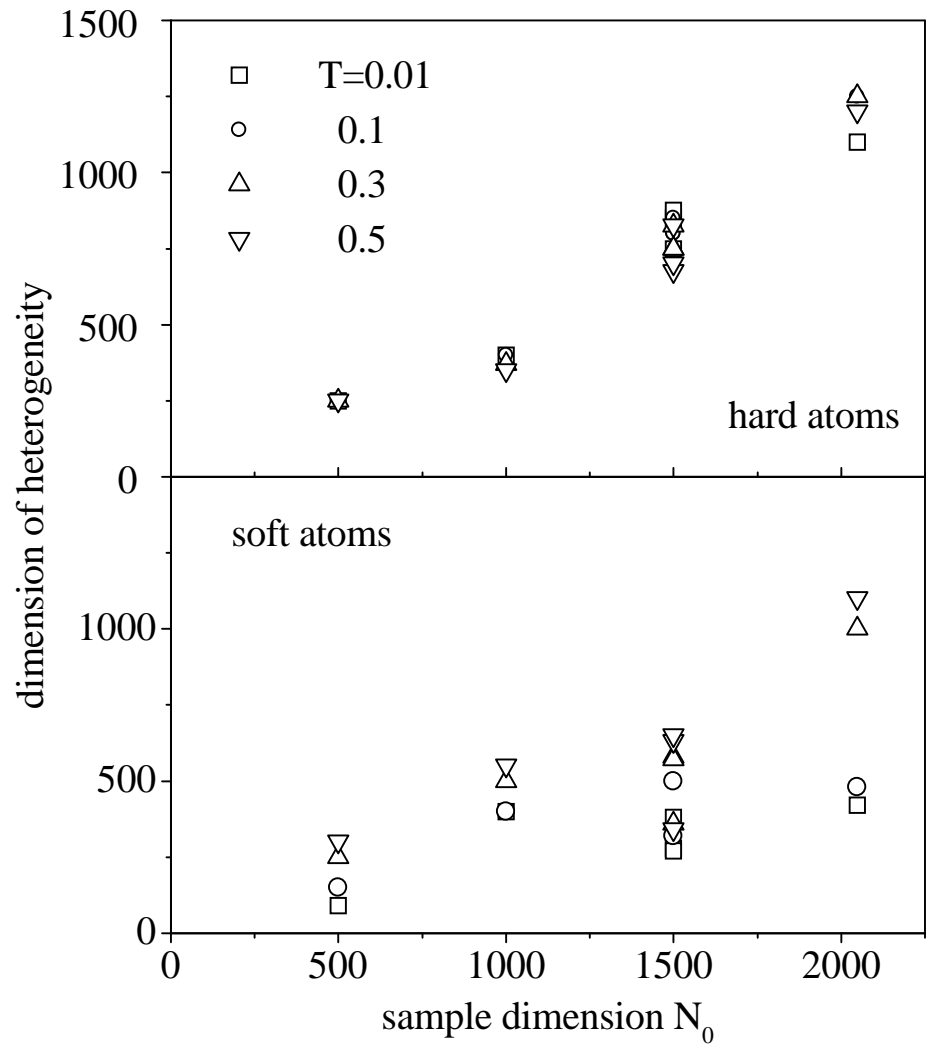
\begin{figure}
\label{f5}
\caption{Dimension of hard (upper) and soft (lower) heterogeneities,
evaluated as
described in the text, at different temperatures, as a function of the total
number of atoms in the sample ($N_0$). For $N_0=1500$ data for three different
samples are reported.}
\end{figure}


\begin{references}

\bibitem{zach32}W. H. Zachariasen, J. Am. Chem. Soc. {\bf 54}, 3841 (1932).

\bibitem{duv90}E. Duval, A. Boukenter and T. Achibat, J. Phys.: Condens.
Matter {\bf 2}, 10 227 (1990).

\bibitem{duv98}E. Duval and A. Mermet, Phys. Rev. B {\bf 58}, 3159 (1998).

\bibitem{Kob97}W. Kob, C. Donati, S. J. Plimpton, P. H. Poole and S. C.
Glotzer, Phys. Rev. lett. {\bf 79}, 2827 (1997).

\bibitem{Poo98}P. H. Poole, C. Donati and S. C. Glotzer, Physica A {\bf
261}, 51 (1998).

\bibitem{Don98}C. Donati, J. F. Douglas, W. Kob, S. J. Plimpton, P. H.
Poole and S. C. Glotzer, Phys. Rev.  Lett. {\bf 80}, 2338 (1998).

\bibitem{Sch91}K. Schmidt-Rohr and H. W. Spiess, Phys. Rev. Lett. {\bf 66},
3020 (1991).

\bibitem{Tra98}U. Tracht, M. Wilhelm, A. Heuer, H. Feng, K. Schmidt-Rohr
and H. W. Spiess, Phys. Rev. lett. {\bf 81}, 2727 (1998).

\bibitem{kus00}T. Kustanovich and Z. Olami, Phys. Rev. B {\bf 61}, 4813
(2000).

\bibitem{Wit01}J. Wittmer, A. Tanguy, J. L. Barrat and L. Lewis, Europhys.
Lett. to be published (2002)

\bibitem{frustra} L. Angelani, M. Montagna, G. Ruocco, and G. Viliani,
Phys. Rev. Lett. {\bf 84}, 4874 (2000).

\bibitem{Bee90}B. W. H. van Beest, G. J. Kramer and R. A. van Santen, Phys.
Rev. Lett. {\bf 64}, 1955 (1990).

\bibitem{Jun01}P. Jund, M. Rarivomanantsoa and R. Julien,  J. Phys. :
Condens. Matter {\bf 12}, 8777 (2000)



\end{references}
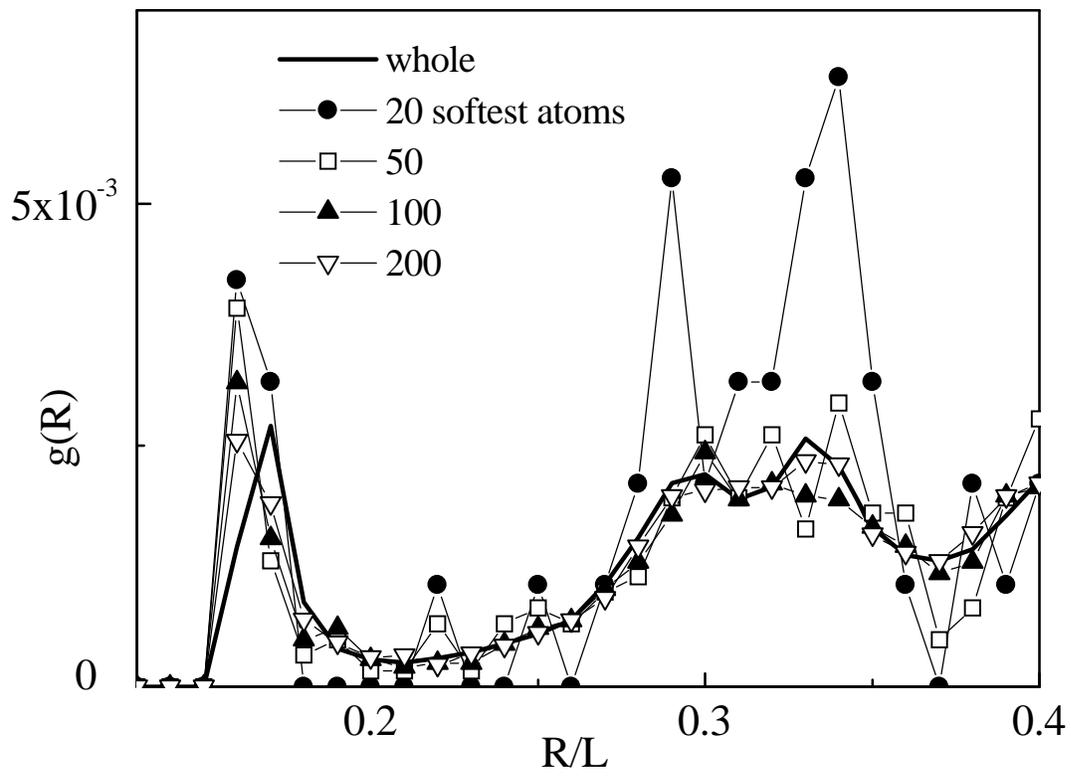
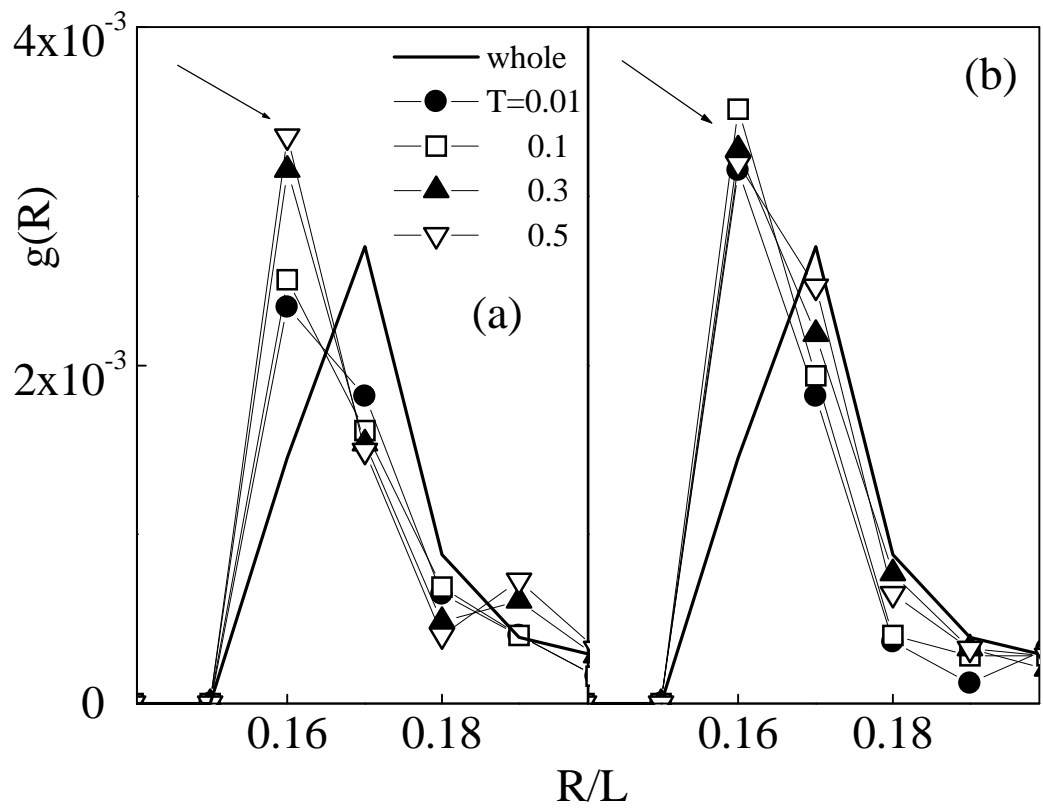
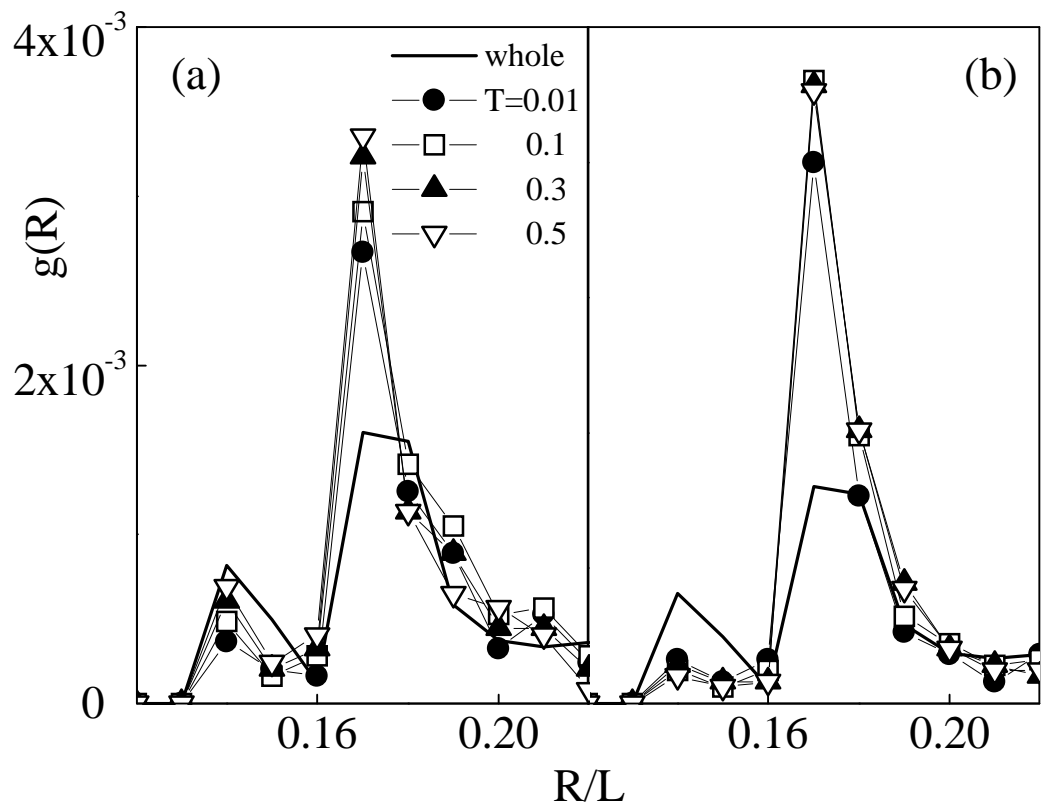
\end{document}